\documentclass[letterpaper,10pt]{article}
\usepackage{osameet2}
\usepackage{graphicx}

\begin{document}

\title{Ultrafast Intersystem Crossing in SO$_2$ and Nucleobases}

\author{Sebastian Mai, Martin Richter, Philipp Marquetand and Leticia Gonz\'alez}
\address{Institute of Theoretical Chemistry, University of Vienna, Währinger Strasse 17, 1090 Vienna, Austria}
\email{leticia.gonzalez@univie.ac.at\\[5mm] 
This manuscript was originally published under
\textit{S. Mai, M. Richter, P. Marquetand \& L. Gonz\'alez Yamanouchi, K.; in Cundiff, S.; de Vivie-Riedle, R.; Kuwata-Gonokami, M. \& DiMauro, L. (Eds.) Ultrafast Intersystem Crossing in SO$_2$ and Nucleobases Ultrafast Phenomena XIX, Springer International Publishing, 2015, 162, 509-513}.
The final publication is available at Springer via http://dx.doi.org/10.1007/978-3-319-13242-6\_124
}

\begin{abstract}
Mixed quantum-classical dynamics simulations show that intersystem crossing between singlet and triplet states in SO$_2$ and in nucleobases takes place on an ultrafast timescale (few 100~fs), directly competing with internal conversion.
\end{abstract}

\section{Introduction}

Transitions involving a change in the spin state of a molecule are formally forbidden in a non-relativistic theoretical frame and are only mediated by spin-orbit couplings (SOCs).
Including SOCs in the description of the excited-state dynamics allows investigating the so-called intersystem crossing (ISC) process\cite{Marian2012WCMS}. Through ISC a singlet state can be transformed radiationlessly into a triplet state, or in general a transition between two states of different multiplicity can occur. ISC is promoted by large SOC and small energy gaps\cite{Marian2012WCMS}. Accordingly, in molecules containing heavy atoms, such as transition metals, where a high density of states and large SOC are present, ultrafast ISC has been observed both experimentally and theoretically, see e.g. Refs.~\cite{Chergui2006ACIE,Tavernelli2011CP,Freitag2014IC}.
However, because SOCs are typically small for organic molecules, ISC is assumed to be one of the slowest forms of relaxation in photochemistry and photophysics, taking place on the order of picoseconds to microseconds\cite{Cadet1990}. Only recently, for a number of molecules including light atoms it has been shown that ISC can indeed occur on an fs timescale and compete with IC\cite{Penfold2012JCP}.

The study of photoinduced reactions can be tackled theoretically in two ways, which are complementary to each other. Solving the time-independent Schr\"odinger equation allows for optimizing critical points on the excited-state potential energy surfaces and proposing pathways that connect these points. A time-dependent approach provides the relevant reaction pathways following the dynamics of the system in real time. An advantage of dynamical simulations is that they also allow to obtain the time constants and quantum yields associated to these pathways.

In this work we study the excited-state dynamics of SO$_2$ and the photophysics of DNA/RNA nucleobases using ab initio molecular dynamics in the form of surface-hopping, according to Tully's fewest switches criterion\cite{Tully1990JCP} within the recently developed SHARC variant\cite{Richter2011JCTC} (Surface Hopping including ARbitrary Couplings), which permits  to describe simultaneously ISC and internal conversion (IC) on the same footing. In both examples, it is illustrated that despite the lack of heavy atoms, ultrafast ISC takes place directly competing with IC.

\section{Methodology}

At least two bases to describe electronic states are distinguished within SHARC. One is the basis of the eigenfunctions of the molecular Coulomb Hamiltonian (MCH) --this is the basis employed in most electronic structure codes. In this representation, coupling between states of the same multiplicity are described by non-adiabatic couplings, while states of different multiplicity are coupled by off-diagonal elements in the Hamiltonian matrix.
Alternatively, another basis in which the Hamiltonian is diagonalized and its eigenfunctions are spin-mixed states coupled only by non-adiabatic couplings is possible. The transformation between the two basis is governed by a unitary matrix $\mathbf{U}$,
\begin{equation}
  \mathbf{H}^{\mathrm{diag}}=\mathbf{U}^\dagger\mathbf{H}^{\mathrm{MCH}}\mathbf{U}
  \quad\quad\mathrm{and}\quad\quad
  \mathbf{c}^{\mathrm{diag}}=\mathbf{U}^\dagger\mathbf{c}^{\mathrm{MCH}},
  \nonumber
\end{equation}
which is also used to transform the gradients and non-adiabatic couplings.

The latter basis is the one employed for surface hopping within the SHARC methodology\cite{Richter2011JCTC,Mai2013CPC}, as it presents a number of advantages.
First, it includes the effect of the SOC on the shape of the potential energy surfaces, which is known as the Zeeman effect.
Second, in this basis all couplings are localized at regions of the potential energy surfaces where the states are close in energy. This implies that trajectories only hop in these localized regions---in the spirit of the fewest-switches approach\cite{Tully1990JCP}.
Third, it correctly accounts for rotational invariance between the multiplet components and thus allows to easily include all of these components.\cite{Granucci2012JCP}

\section{Results}

In the following, the photophysics of two different molecular systems will be discussed: The SO$_2$ and DNA/RNA nucleobases, especially focusing on the two main tautomers of cytosine.

\subsection{Dynamics of SO$_2$}

\begin{figure}[h!]
  \centering
  \includegraphics[scale=0.8]{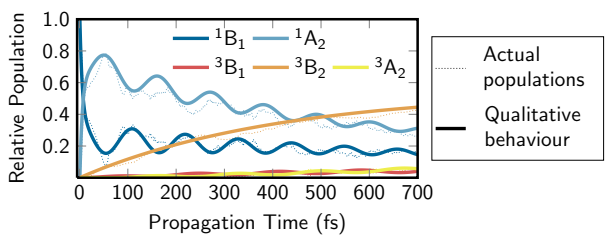}
  \caption{Time evolution of the excited-state populations in SO$_2$. The main processes are periodical IC between $^1B_1$ and $^1A_2$ and ISC towards $^3B_2$.}
  \label{fig:so2}
\end{figure}

In SO$_2$ a  complicated vibrational structure, arising from IC between the $^1A_2$ and $^1B_1$ singlet states, has been observed in the first allowed band of the absorption spectrum.
Another band of very weak intensity exists at energies slightly lower than the first allowed band, which shows an appreciable Zeeman effect\cite{Douglas1958CJP} and hence is proposed to arise from ISC.
Modern quantum chemistry indicates the presence of three triplet states ($^3A_2$, $^3B_1$, $^3B_2$) in this energetic region of the spectrum, but neither experiments nor static ab initio calculations could elucidate the relative importance of these states.

The results of SHARC dynamics simulations\cite{Mai2014_SO2} on SO$_2$ based on on-the-fly MRCI wavefunctions are shown in Figure~\ref{fig:so2}. Starting from the spectroscopically bright $^1B_1$ state, oscillatory population transfer by IC between $^1B_1$ and $^1A_2$ occurs. Additionally, the $^3B_2$ triplet state is populated (mainly from the $^1A_2$ state) on a 410~fs timescale, showing that this state is strongly participating in the excited-state dynamics. Already after 700~fs, more than 40\% of the population has crossed to the $^3B_2$ state. On the contrary, the triplet states $^3A_2$ and $^3B_1$ do not show a large participation in the dynamics.
These findings agree well with recent experimental studies based on the TRPEPICO (time-resolved photo-electron photo-ion coincidence spectroscopy) method\cite{Wilkinson2014JCP} as well as with recent quantum dynamical simulations~\cite{Leveque2014JCP}.

\subsection{Dynamics of DNA Nucleobases}

\begin{figure}[h!]
  \centering
  \includegraphics[scale=0.8]{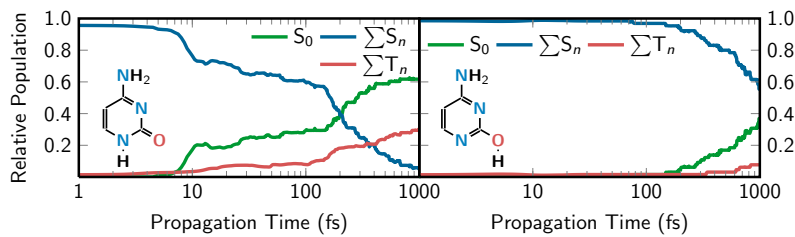}
  \caption{Populations of the electronic states in cytosine tautomers (left: keto, right: enol). The keto form shows fast ground state relaxation and ISC, while the enol shows only much slower relaxation.}
  \label{fig:cyt}
\end{figure}

A number of experiments have been performed on isolated nucleobases in gas phase in order to understand their excited-state dynamics. Cytosine is a particularly complex case since it involves several stable tautomers in the gas phase and the presence of multiple deactivation pathways~\cite{Gonzalez-Vazquez2010C}.
As Figure~\ref{fig:cyt} shows, the time evolution of the excited-state populations of the keto and enol tautomers are significantly different~\cite{Mai2013CPC}.
In the keto tautomer, already after 10~fs, a notable fraction of the population is transferred to the electronic ground state and after 1 ps more than 60\% of the population is non-radiatively relaxed to the ground state. Additionally, almost 30\% of the population undergo ISC to the triplet manifold during this time.
On the contrary, in the enol tautomer, ground state relaxation is much slower. After the first ps, still about 60\% of the population remain in the excited singlet states. Moreover, the enol form shows almost no ISC to the triplet manifold. According to our calculations, the  experimentally observed multiexponential decay of cytosine\cite{Ho2011JPCA} could be then explained by the presence of both the keto and enol tautomer with their significantly different relaxation timescales.
The analysis of our trajectories~\cite{Mai2013CPC} reveals that the keto tautomer relaxes to the ground state mainly through a three-state conical intersection, which is planar and thus quickly accessible from the Franck-Condon region (hence explaining relaxation to the ground state starts already after 10~fs). In the enol tautomer, ground state relaxation involves conical intersections with highly distorted ring structures, necessitating large-scale motion of the ring atoms. Additionally, it was found that ISC in the keto tautomer is fast, compared to the enol tautomer, due to the presence of the carbonyl group that increases SOC substantially.

Analogous SHARC simulations on thymine and uracil~\cite{Richter2014PCCP} reveal that ISC also takes places on an ultrafast time scale. Especially in thymine, ISC is important since it is proposed to lead to thymine-thymine dimerization, one of the most prevalent types of DNA photodamage.

\section{Conclusion}

This contribution shows that the SHARC methodology is an efficient tool to study the excited-state dynamics of molecules in the presence of non-adiabatic and spin-orbit couplings, providing significant insight into the processes occurring on a molecular level during the first few ps.

% \bibliographystyle{osajnl}
% \bibliography{allrefs}

\end{document}